\documentclass[]{spie}  %>>> use for US letter paper
\usepackage{amsmath,amsfonts,amssymb}
\usepackage{graphicx}
\usepackage{booktabs}
\usepackage[colorlinks=true, allcolors=blue]{hyperref}

\title{Optical Characterization of the BICEP Array 150 and 220/270 GHz CMB Polarimeters in the 2026 Season}

\author[a]{M.~Izquierdo~Poza}
\author[b]{P.~A.~R.~Ade}
\author[c,d]{Z.~Ahmed}
\author[e]{M.~Amiri}
\author[f]{D.~Barkats}
\author[g]{R.~Basu~Thakur}
\author[h]{C.~A.~Bischoff}
\author[i]{D.~Beck}
\author[g,j]{J.~J.~Bock}
\author[k]{V.~Buza}
\author[d,i]{B.~Cantrall}
\author[g]{J.~R.~Cheshire~IV}
\author[l]{J.~Connors}
\author[m]{J.~Cornelison}
\author[n]{M.~Crumrine}
\author[g]{A.~J.~Cukierman}
\author[l]{E.~Denison}
\author[o]{L.~Duband}
\author[f]{M.~Echter}
\author[p]{M.~Eiben}
\author[f,q]{B.~D.~Elwood}
\author[g]{S.~Fatigoni}
\author[r]{J.~P.~Filippini}
\author[i]{A.~Fortes}
\author[g]{M.~Gao}
\author[h]{C.~Giannakopoulos}
\author[h]{N.~Goeckner-Wald}
\author[h]{D.~C.~Goldfinger}
\author[h]{J.~A.~Grayson}
\author[g]{A.~Greathouse}
\author[f]{P.~K.~Grimes}
\author[e]{M.~Halpern}
\author[c,d]{S.~Henderson}
\author[n]{T.~D.~Hoang}
\author[l]{J.~Hubmayr}
\author[g]{H.~Hui}
\author[i]{K.~D.~Irwin}
\author[g]{J.~H.~Kang}
\author[a]{K.~S.~Karkare}
\author[g]{S.~Kefeli}
\author[f,q]{J.~M.~Kovac}
\author[i]{C.~Kuo}
\author[n,s]{K.~Lasko}
\author[g]{K.~Lau}
\author[h]{M.~Lautzenhiser}
\author[i]{T.~Liu}
\author[k,r]{S.~C.~Mackey}
\author[n]{N.~Maher}
\author[j]{K.~G.~Megerian}
\author[g]{L.~Minutolo}
\author[g]{L.~Moncelsi}
\author[i]{Y.~Nakato}
\author[g,j]{H.~T.~Nguyen}
\author[g,j]{R.~O'Brient}
\author[f]{S.~N.~Paine}
\author[g]{A.~Patel}
\author[f]{M.~A.~Petroff}
\author[f,q]{A.~R.~Polish}
\author[o]{T.~Prouve}
\author[n]{C.~Pryke}
\author[l]{C.~D.~Reintsema}
\author[g]{T.~Romand}
\author[i]{M.~Salatino}
\author[g]{A.~Schillaci}
\author[f]{B.~Schmitt}
\author[n,s]{B.~Singari}
\author[g,j]{A.~Soliman}
\author[f]{T.~St.~Germaine}
\author[g]{A.~Steiger}
\author[g]{B.~Steinbach}
\author[b]{R.~Sudiwala}
\author[d,i]{K.~L.~Thompson}
\author[b]{C.~Tucker}
\author[j]{A.~D.~Turner}
\author[t]{C.~Verg\`{e}s}
\author[k,r]{A.~G.~Vieregg}
\author[g]{A.~Wandui}
\author[j]{A.~C.~Weber}
\author[n]{J.~Willmert}
\author[c,d]{W.~L.~K,~Wu}
\author[i]{H.~Yang}
\author[k,m]{C.~Yu}
\author[f]{L.~Zeng}
\author[d]{C.~Zhang}
\author[g]{S.~Zhang}

\affil[a]{Department of Physics, Boston University, Boston, MA 02215, USA}
\affil[b]{School of Physics and Astronomy, Cardiff University, Cardiff, CF24 3AA, United Kingdom}
\affil[c]{SLAC National Accelerator Laboratory, Menlo Park, CA 94025, USA}
\affil[d]{Kavli Institute for Particle Astrophysics and Cosmology, Stanford University, Stanford, CA 94305, USA}
\affil[e]{Department of Physics and Astronomy, University of British Columbia, Vancouver, British Columbia, V6T 1Z1, Canada}
\affil[f]{Center for Astrophysics, Harvard \& Smithsonian, Cambridge, MA 02138, USA}
\affil[g]{Department of Physics, California Institute of Technology, Pasadena, CA 91125, USA}
\affil[h]{Department of Physics, University of Cincinnati, Cincinnati, OH 45221, USA}
\affil[i]{Department of Physics, Stanford University, Stanford, CA 94305, USA}
\affil[j]{Jet Propulsion Laboratory, California Institute of Technology, Pasadena, CA 91109, USA}
\affil[k]{Kavli Institute for Cosmological Physics, University of Chicago, Chicago, IL 60637, USA}
\affil[l]{National Institute of Standards and Technology, Boulder, CO 80305, USA}
\affil[m]{Argonne National Laboratory, High Energy Physics Division, Lemont, IL 60439, USA}
\affil[n]{School of Physics and Astronomy, University of Minnesota, Minneapolis, MN 55455, USA}
\affil[o]{Service des Basses Temperatures, Commissariat a l’Energie Atomique, 38054 Grenoble, France}
\affil[p]{Science Institute, University of Iceland, Dunhagi 5, 107 Reykjavik, Iceland}
\affil[q]{Department of Physics, Harvard University, Cambridge, MA 02138, USA}
\affil[r]{Department of Physics, Grainger College of Engineering, University of Illinois Urbana-Champaign, Urbana, IL 61801, USA}
\affil[s]{Minnesota Institute for Astrophysics, University of Minnesota, Minneapolis, MN 55455, USA}
\affil[t]{Physics Division, Lawrence Berkeley National Laboratory, Berkeley, CA 94720, USA}

\authorinfo{Send correspondence to M. Izquierdo Poza.
E-mail: marcelai@bu.edu}

\begin{document} 
\maketitle
\begin{abstract}
BICEP Array (BA) is the current-generation instrument in the BICEP series of small-aperture, on-axis refracting telescopes at the South Pole, designed to constrain the tensor-to-scalar ratio $r$ through degree-scale measurements of B-mode polarization in the cosmic microwave background (CMB). As BA pushes to deeper sensitivity, control of instrumental systematics, and beam shape mismatch between the co-located orthogonally polarized detectors in particular, has become an increasingly important factor in translating raw sensitivity into a robust constraint on $r$. In these proceedings we report on the 2026 far field beam mapping (FFBM) campaign, which used a thermal chopped source to characterize the BA2 (150~GHz) and BA3 (220/270~GHz) beams. We fit two-dimensional elliptical Gaussians to each detector's beam, derive per-pair differential parameters (differential pointing, beamwidth, and ellipticity). The resulting high-signal-to-noise array-averaged beam maps are used to compute the beam window function $B_l$ for the power spectrum analysis, while the individual per-detector beams feed dedicated beam convolution simulations used to validate the temperature-to-polarization (T$\rightarrow$ P) deprojection procedure. After correcting for the chopper aperture, the recovered beamwidths follow the expected $\lambda/D$ ordering. We also describe two pipeline improvements carried out during the 2026 campaign:\ an out-and-back jackknife for noise quantification and an elnod-based gain calibration.
\end{abstract}


\section{INTRODUCTION}
\label{sec:intro}  % \label{} allows reference to this section
%\begin{itemize}
%    \item Science motivation, constraining r and the role of beam systematics.
%   \item T→P leakage and deprojection framework
%   \item Overview of BA2/BA3 receivers and the 2026 observing season context (new detector tiles)
%\end{itemize}
The clearest signature of cosmic inflation, a proposed burst of exponential expansion an instant after the Big Bang, is a particular pattern in the polarization of the cosmic microwave background (CMB). Inflation explains several features of the observed universe, including its near-flatness and monopole problems of standard cosmology, while also explaining the origin of structure by stretching quantum fluctuations to macroscopic scales. \cite{planck2015inflation} Inflation also predicts two kinds of perturbations to the metric in both scalars (density waves) and tensors (gravitational waves). At the surface of the last scattering, scalar perturbations can only produce a curl-free (E-mode) pattern in the CMB polarization, while tensor perturbations also produce a curl-type (B-mode) pattern. \cite{polnarev1985} Aside from the well-understood B-mode contribution from the gravitational lensing of E-modes, no other primordial process is expected to generate B-mode power at degree angular scales. A detection above this lensing floor would therefore be direct evidence for inflation, and its amplitude---the tensor-to-scalar ratio $r$---would fix the energy scale at which inflation occurred. 

The BICEP experiments have pursued this measurement from the Amundsen--Scott South Pole Station since 2006, observing a sky patch corresponding to $1 \%$ of the sky chosen for its low polarized foreground emission. \cite{bicep2_II_2014} Because the B-mode signal from inflation is expected to peak near $l \sim 80$, each generation of the instrument has used a small, fast, on-axis refracting telescope, prioritizing systematics control. The most recent published result, using data through the 2018 observing season (BK18), constrains $r<$0.036 at 95$\%$ confidence, the tightest constraint to date on primordial gravitational waves \cite{bk18_2021}. As the statistical sensitivity of the BICEP program has grown with each new receiver generation, control of instrumental systematics and of beam-shape mismatch between pair detectors in particular, is increasingly important in translating raw sensitivity into a robust constraint on $r$, motivating the dedicated beam-characterization effort described here. %In 2014 BICEP2 reported a detection of B-mode power at degree angular scales at $150$ GHz, \cite{bicep2_2014_detection} which was subsequently confirmed by the Keck Array. \cite{bicepkeck_V_2015} Because both thermal dust emission at high frequencies and synchrotron emission at low frequencies can mimic primordial B-mode signal, disentangling the two requires observations at multiple frequencies. Combining BICEP2/Keck Array data with Planck and WMAP maps showed that the original excess was consistent with Galactic dust and tightened the constraint to $r < 0.07$ at $95\%$ confidence. \cite{joint_planck_2015,improved_95ghz_2016} Subsequently we added dedicated $220$ GHz dust-monitoring receivers to the Keck Array and built BICEP3, a larger-aperture, faster  $95$ GHz receiver, to push deeper at a frequency with minimal foreground contamination. \cite{karkare2016bicep3,wu2016bicep3perf} As the statistical sensitivity of the BICEP program has grown with each new receiver generation, control of instrumental systematics and of beam-shape mismatch between pair detectors in particular, is increasingly important in translating raw sensitivity into a robust constraint on $r$, motivating the dedicated beam-characterization effort described here. 

BICEP's optical design detects polarization from the difference between two co-located detectors sensitive to orthogonal polarization states. \cite{bicep2_II_2014} Because this differencing strategy relies on two detectors having identical beams, any discrepancy between their beam shapes couples a portion of the much brighter unpolarized CMB signal into the polarization difference. This temperature-to-polarization ($T\rightarrow P$) leakage is a leading instrumental systematic for pair-differencing CMB experiments, capable of biasing $r$ if it goes uncorrected. \cite{bicepkeck_IV_2015} Our first level of mitigation is deprojection: %rather than being built from the measured beam mismatch itself, deprojection templates are constructed from spatial derivatives of the measured beam mismatch itself, deprojection 
templates of the Planck temperature sky and its first and second derivatives are constructed, with different combinations of the templates corresponding roughly to lowest-order beam mismatch modes. %are constructed from spatial derivatives of the Plank CMB temperature map,\cite{bicep2_II_2014} with each derivative order corresponding to a different beam mismatch mode. 
The timestreams are regressed against these templates, and the best fit template is subtracted, removing leakage from that particular mode without relying on external knowledge of the true beam mismatch. \cite{bicep2_III_2015} Independent, high-fidelity measurements of each detector's beam---given by the far field beam maps (FFBM) described here---provide a critical cross-check that the modes removed by deprojection correspond to real beam mismatch, and allow us to calculate any residual leakage that is not removed by deprojection of the lowest-order beam mismatch modes.\cite{karkare2016bicep3,bk15xi_2020,stgermaine2020}

BICEP Array (BA) is the current generation of the program, a four-receiver upgrade that replaces the smaller-aperture Keck Array receivers with larger, more sensitive telescopes at 30/40, 150, and 220/270~GHz, while one Keck receiver (K5, 270~GHz) currently remains in the field during the transition. %BA2, the 150~GHz receiver, serves as the program's primary low-foreground CMB science channel and  had a new low-pass edge filter installed between 2024 and 2026 observing seasons. 
BA3, the newest receiver at 220/270~GHz, entered the 2026 season with seven previously unmeasured detector tiles. %, and checking the general receiver health was a focus of the 2026 FFBM campaign. %By combining a low-foreground 150~GHz science channel with dedicated low and high frequency dust and synchrotron monitors on a single mount, BICEP Array is designed to push well past current constraints on $r$. 
These proceedings report on the 2026 BA FFBM campaign, which prioritized health and performance checks for the new BA3 (220/270~GHz) receiver and extended BA2 (150~GHz) coverage to newly installed tiles, while providing a cross-season consistency check against the 2024 BA2 FFBM dataset---a low-pass edge filter was installed for the 2025 season. Section 2 describes the far field beam mapping setup and the 2026 dataset, including the receiver configuration for that season. Section 3 presents the two-dimensional elliptical Gaussian fits used to derive per-detector and per-pair beam parameters, along with array-level summary statistics for BA2 and BA3. Section 4 describes the composite and array-averaged beam maps. Section 5 describes two pipeline improvements developed during the 2026 campaign, and Section 6 concludes and outlines the next steps.

\section{FAR FIELD BEAM MEASUREMENT SETUP}

\subsection{Flat Mirror and Thermal Chopped Source}
\label{sec:title}

\begin{figure}%[htbp]
    \centering
    \includegraphics[width=0.9\linewidth]{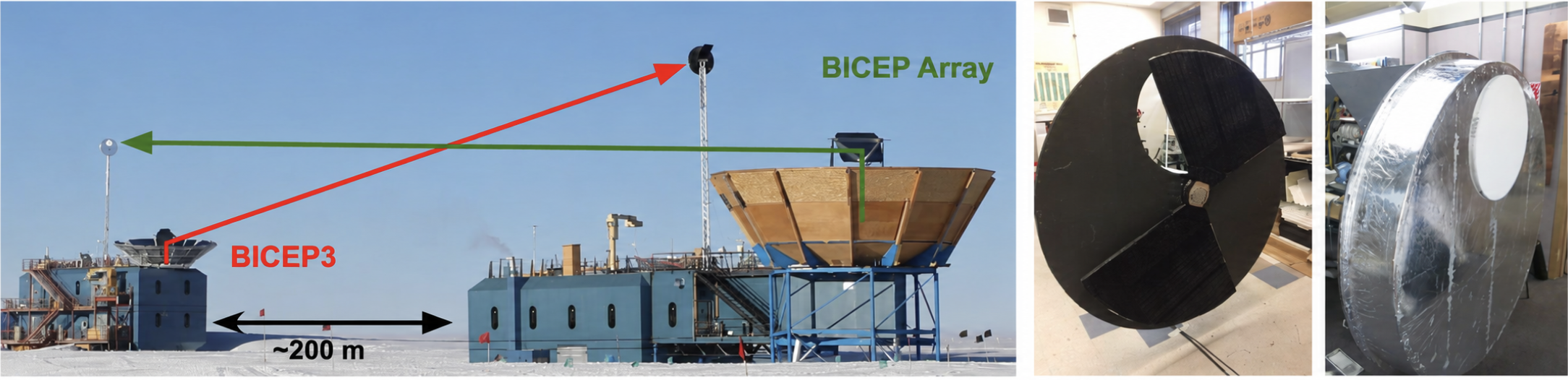}
    \caption{Left: Image of Keck Array and BICEP3 in Feb 2017 performing simultaneous beam mapping campaigns, with redirecting mirrors and thermal sources visible. Middle: The thermal chopper undergoing lab tests with the enclosure off. Right: The same chopper with enclosure on, sealed with Zotefoams HD30.\cite{bk15xi_2020}}
    \label{fig:FFBMsetup}
\end{figure}

Beams are mapped using a thermal chopped source located at a distance of $\approx 211$~m, reflected off a steerable flat mirror attached on the telescope mount (see Figure~\ref{fig:FFBMsetup}), following the approach developed for BICEP2, Keck Array, and BICEP3.\cite{bicepkeck_IV_2015} The telescope field of view is scanned across the artificial source at a fixed boresight rotation angle (dk) to sample all detector beams in that orientation; the scan is then repeated at multiple boresight rotation angles and combined to build composite, array-averaged beam maps (Section 4). 

\subsection{2026 Dataset}
%\begin{itemize}
 %   \item Schedule design and coverage
  %  \item BA3-220/270 and BA2-150 observations
%\end{itemize}

\begin{figure}%[htbp]
    \centering
    \includegraphics[width=0.9\linewidth]{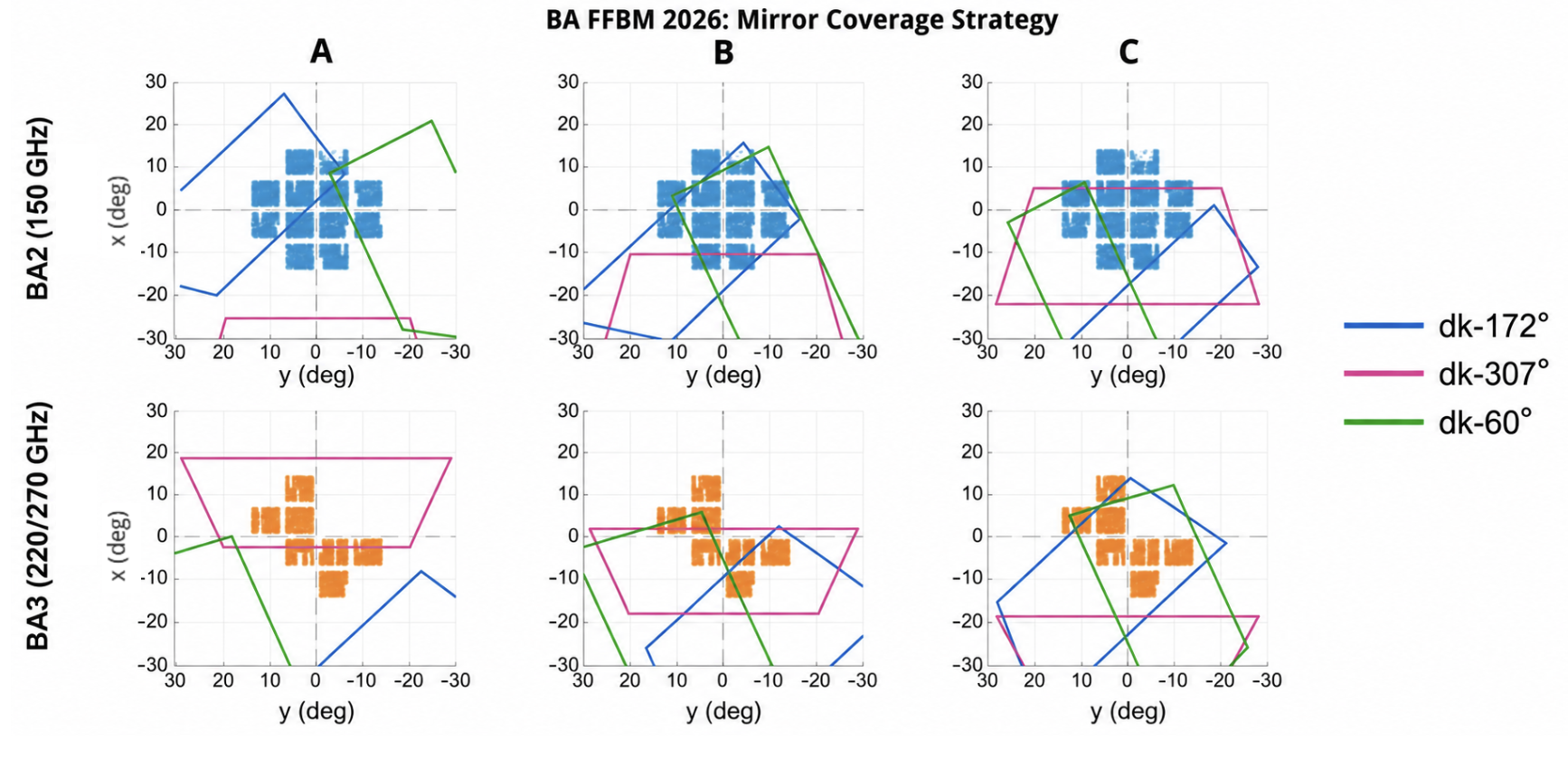}
    \caption{Footprint of the mirror overlaid on beams at different boresight angles, illustrating the mirror coverage strategy needed to fully sample both receivers across the focal plane using three different dk (boresight angles). The columns A, B, and C correspond to three different mirror positions. BA2 is shown in blue and BA3 in orange.}
    \label{fig:mirror}
\end{figure}

Beam mapping schedules for the 2026 campaign were designed for full coverage of BA2 and BA3 at multiple boresight rotations. The main challenge of the setup is that the flat mirror is not large enough to fully cover all
detector tiles in the newly-installed receivers. For this reason, each far-field beam-mapping campaign
must be carefully planned in advance, using a studied combination of boresight angles and mirror
positions to cover all detectors of interest, as shown in Figure~\ref{fig:mirror}. Table~\ref{tab:receivers} summarizes the four receivers currently in BICEP Array. BA3 (220/270~GHz) was the primary target of the campaign given its seven detector tiles that have never been beam mapped before; BA2 (150~GHz) was mapped for cross-season consistency against the 2024 dataset (Section 3.3)  with a focus on two tiles that were not mapped in previous seasons. 

\setlength{\floatsep}{12pt plus 2pt minus 2pt}

\begin{table}%[htbp]
\centering
\caption{BICEP Array and Keck receivers observed in the 2026 FFBM campaign.}
\label{tab:receivers}
\begin{tabular}{lll}
\toprule
\textbf{Receiver} & \textbf{GHz} & \textbf{Notes} \\
\midrule
BA1 & 30 / 40   & Synchrotron channel, previously mapped \\
BA2 & 150       & Cross-season check with new edge filter and 2 new tiles \\
BA3 & 220 / 270 & New receiver with 7 tiles \\
K5  & 270       & Keck receiver, smaller aperture \\
\bottomrule
\end{tabular}
\end{table}

\section{GAUSSIAN BEAM PARAMETERS}
\label{sec:sections}
 
\subsection{Coordinate System and 2D Gaussian Fits}

Following the convention established in previous analyses\cite{karkare2016bicep3}, each beam is projected into a local ($x', y'$) coordinate system defined at that detector's own position in the focal plane. For each optically paired A/B detector, we fit a two-dimensional elliptical Gaussian of the form
\begin{equation}
B(\mathbf{x}) = \frac{1}{\Omega}\exp\!\left[-\frac{1}{2}(\mathbf{x}-\boldsymbol{\mu})^{T}\Sigma^{-1}(\mathbf{x}-\boldsymbol{\mu})\right],
\label{eq:gaussian}
\end{equation}
with beam center $\boldsymbol{\mu} = (x,y)$ and covariance
\begin{equation}
\Sigma =
\begin{pmatrix}
\sigma^2(1+p) & c\sigma^2 \\
c\sigma^2 & \sigma^2(1-p)
\end{pmatrix},
\label{eq:covariance}
\end{equation}
where $\sigma$ is the beamwidth and $p$ and $c$ parametrize the ellipticity along the focal-plane axes and
their diagonal.

\begin{figure}%[htbp]
    \centering
    \includegraphics[width=0.7\linewidth]{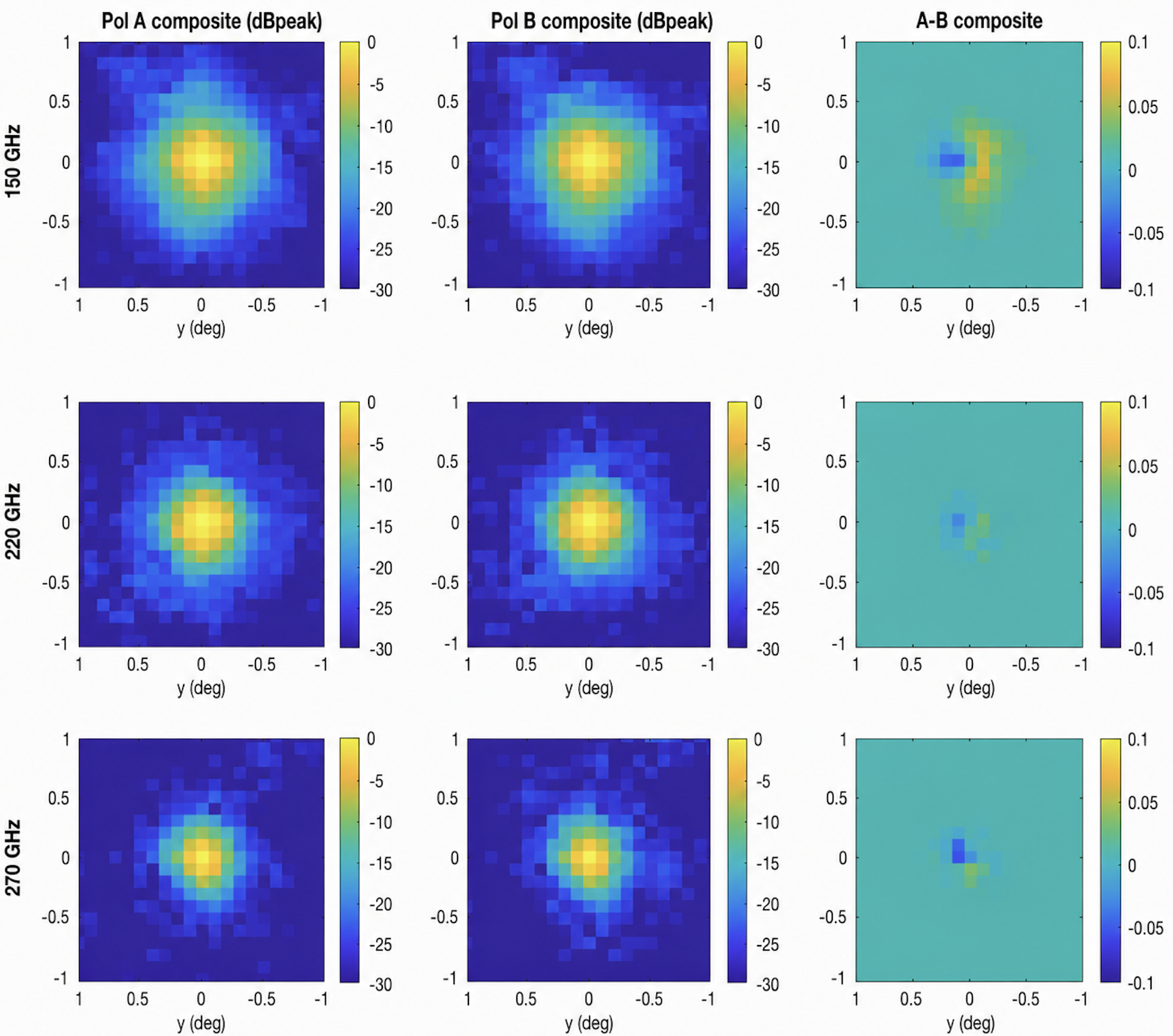}
    \caption{Example per-detector beam maps at 150~GHz (top) and 220~GHz (middle), and 270~GHz (bottom) for pol\,A, pol\,B, and the A$-$B difference. The broadening due to chopper aperture convolution is more pronounced at 220 and 270~GHz}
    \label{fig:bands}
\end{figure}

Figure~\ref{fig:bands} shows example per-detector beam maps at 150, 220, and 270~GHz for A and B polarizations of a pair, together with the A-B difference map. Each panel is a composite built from two individual beam map measurements of the same detector. The beams at all three frequencies are broadened by convolution of the chopper source, an effect that is more pronounced at 220 and 270~GHz (Section 3.3). 

\subsection{Per-Pair Parameters}

The fitted parameters for each A/B pair are used to define the per-pair differential parameters to cross-check the  $T\rightarrow P$ deprojection coefficients.
\begin{align}
dx &= x_A - x_B, \quad dy = y_A - y_B \qquad \text{(differential pointing)} \\
d\sigma &= \sigma_A - \sigma_B \qquad\qquad\qquad\quad\ \text{(differential beamwidth)} \\
dp &= p_A - p_B, \quad dc = c_A - c_B \qquad \text{(differential ellipticity)}
\end{align}
Differential gain between A and B is expected to differ from the CMB deprojection coefficients since beam maps are taken on the detectors' aluminum transition instead of the lower-loading titanium transition. Similarly, the differential ellipticity deprojection coefficient is expected to differ from the beam map derived value, since the CMB fit also absorbs the real, known $\Lambda $CDM $TE$ correlation, which couples to the same ellipticity mode template. 

\begin{figure}%[htbp]
    \centering
    \includegraphics[width=0.6\linewidth]{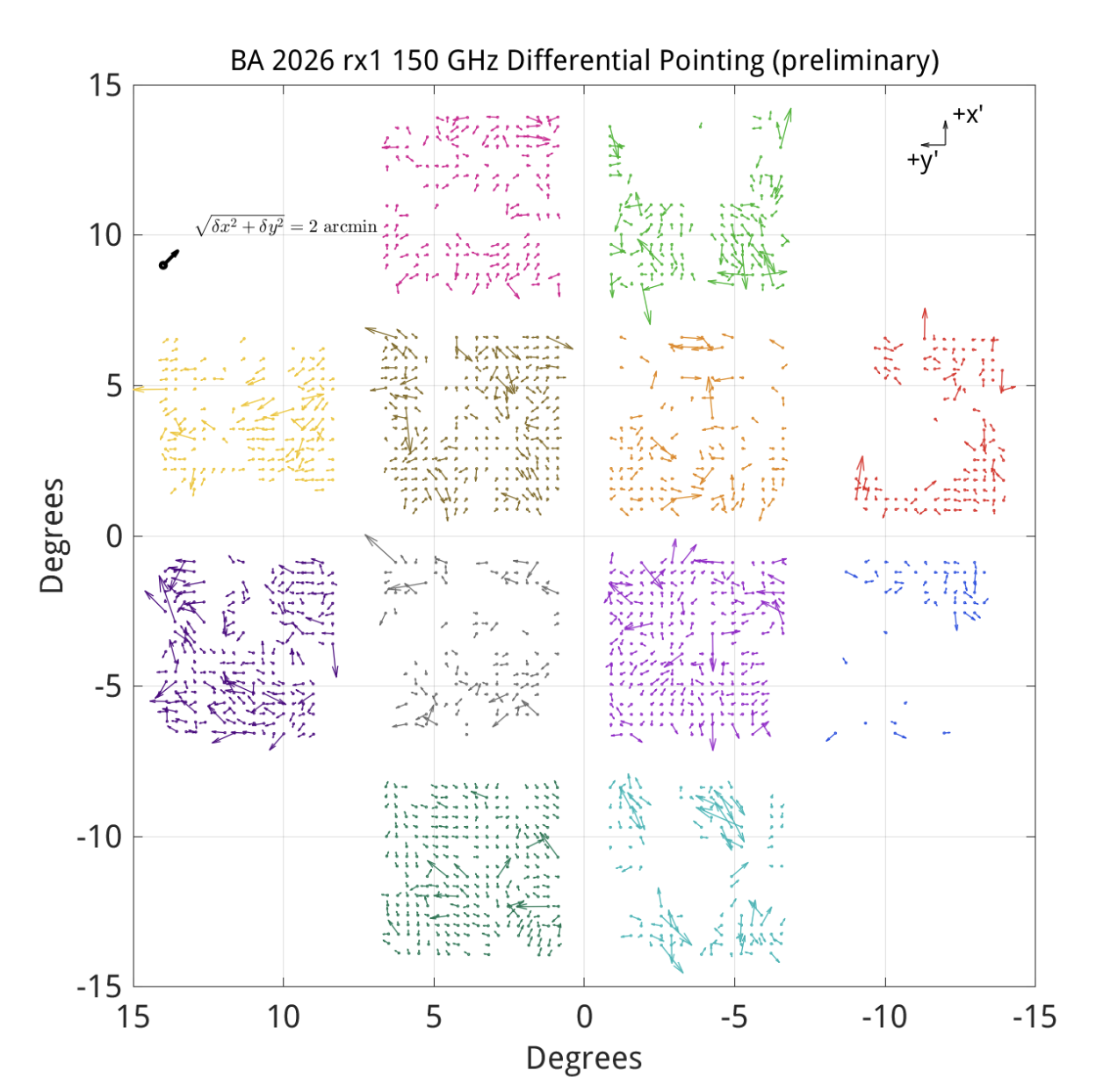}
    \caption{Focal-plane map of differential pointing $|\Delta\mathbf{p}|$ for BA2-150 GHz; different color per tile. Arrows point from the A detector location to the B detector location, so both the magnitude $|\Delta p| = \sqrt{dx^2+dy^2}$ (arrow length) and direction of the A-to-B offset are shown. Note that the length of the arrows should not be compared to the x/y axes.}
    \label{fig:diffpoint}
\end{figure}

Figure~\ref{fig:diffpoint} shows the focal-plane distribution of the differential pointing for the BA2 150~GHz receiver. Each arrow points from the A detector location to the B detector location, so arrow length encodes the magnitude $|\Delta p| = \sqrt{dx^2+dy^2}$ while arrow direction shows the orientation of the A-to-B offset, colored by tile.

\subsection{Summary Statistics: BA2 and BA3}

\begin{figure}%[htbp]
    \centering
    \includegraphics[width=0.9\linewidth]{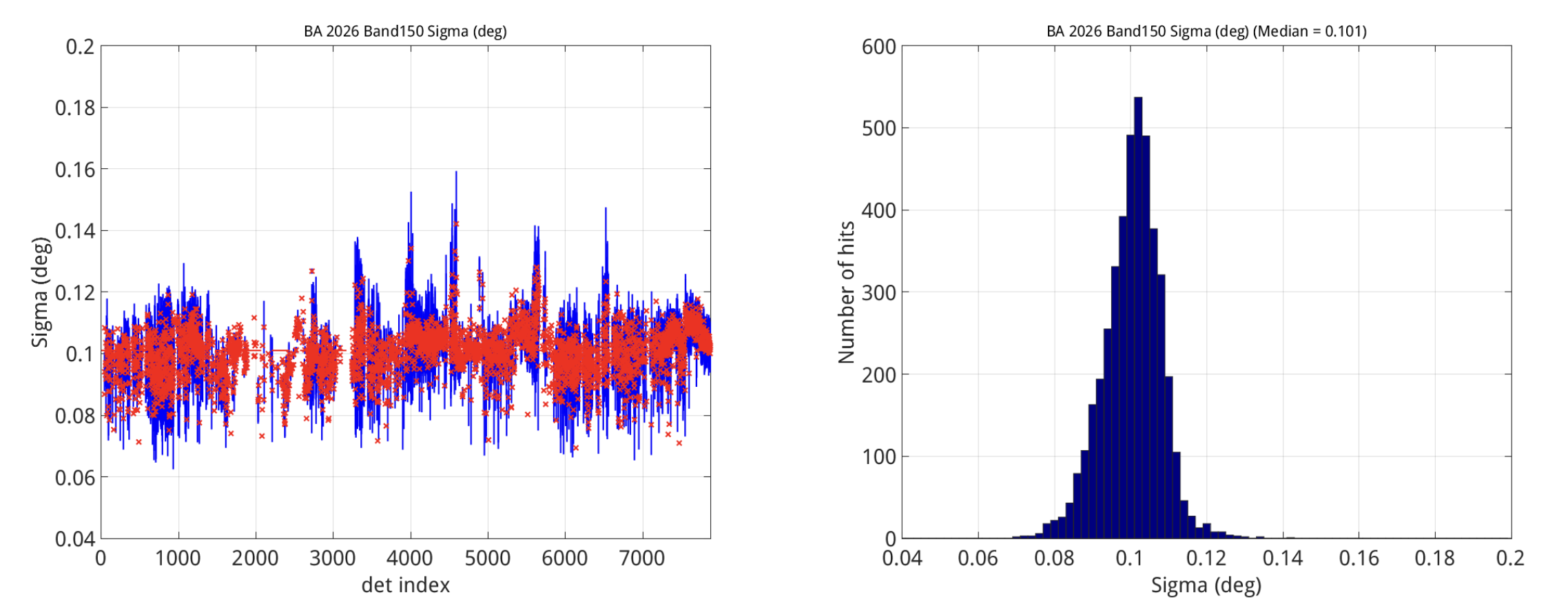}
    \caption{BA2-150 beamwidth $\sigma$. Left: per-detector median $\sigma$ (red crosses) vs.\ detector index, with error bars (blue bars) given by half the 84th--16th percentile spread. Right: histogram of median $\sigma$ across the array.}
    \label{fig:hist}
\end{figure}

Figure~\ref{fig:hist} shows the per-detector beamwidths $\sigma$ for BA2 at 150~GHz. The left panel plots the median $\sigma$ (across all measurements) for each detector against detector index, with error bars given by half the 84th-16th percentile spread of the per-schedule measurement; the right panel shows the corresponding histogram of $\sigma$ across the array. After deconvolving the finite chopper aperture, the corrected beamwidths follow the expected $\lambda/D$ ordering across the bands mapped in 2026, $\sigma_{150} > \sigma_{220} > \sigma_{270}$.

\begin{figure}%[htbp]
    \centering
    \includegraphics[width=0.9\linewidth]{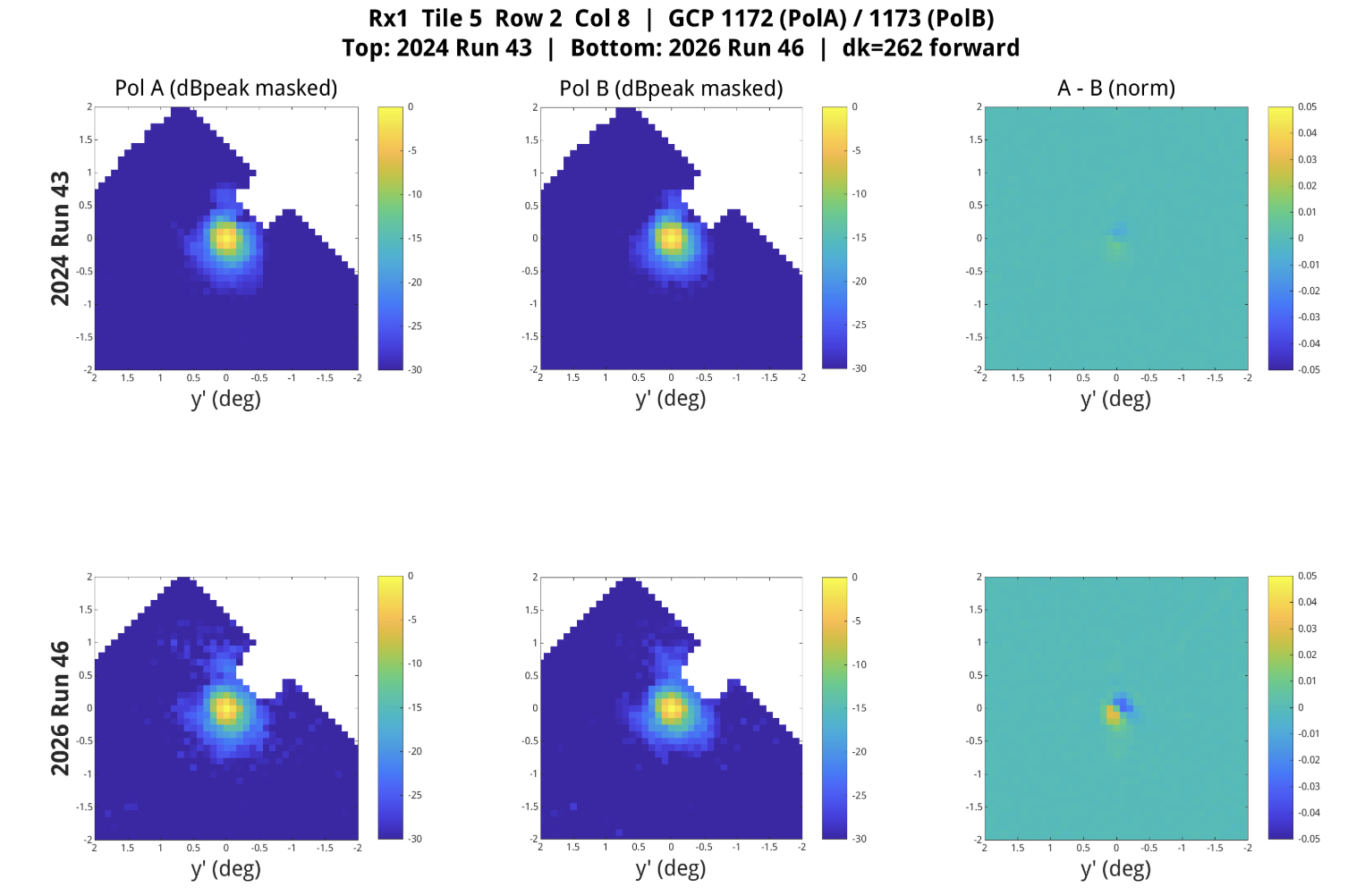}
    \caption{BA2-150 comparison between FFBM 2024 vs. 2026. The source mast and the ground are masked out in each of the component maps.}
    \label{fig:2024vs2026}
\end{figure}

For the 2025 season a low-pass edge filter was added to BA2. To check measurable optical differences introduced by this change, we compared maps of the same detectors before and after the filter was installed, matching mirror position and boresight angle between the two epochs. Figure~\ref{fig:2024vs2026} shows an example for one detector pair for this BA2-150 comparison between the 2024 and 2026 FFBM campaigns, showing similar but not identical beam structure.

\section{COMPOSITE BEAM MAPS AND ARRAY-AVERAGED MAPS}

\begin{figure}%[htbp]
    \centering
    \includegraphics[width=0.75\linewidth]{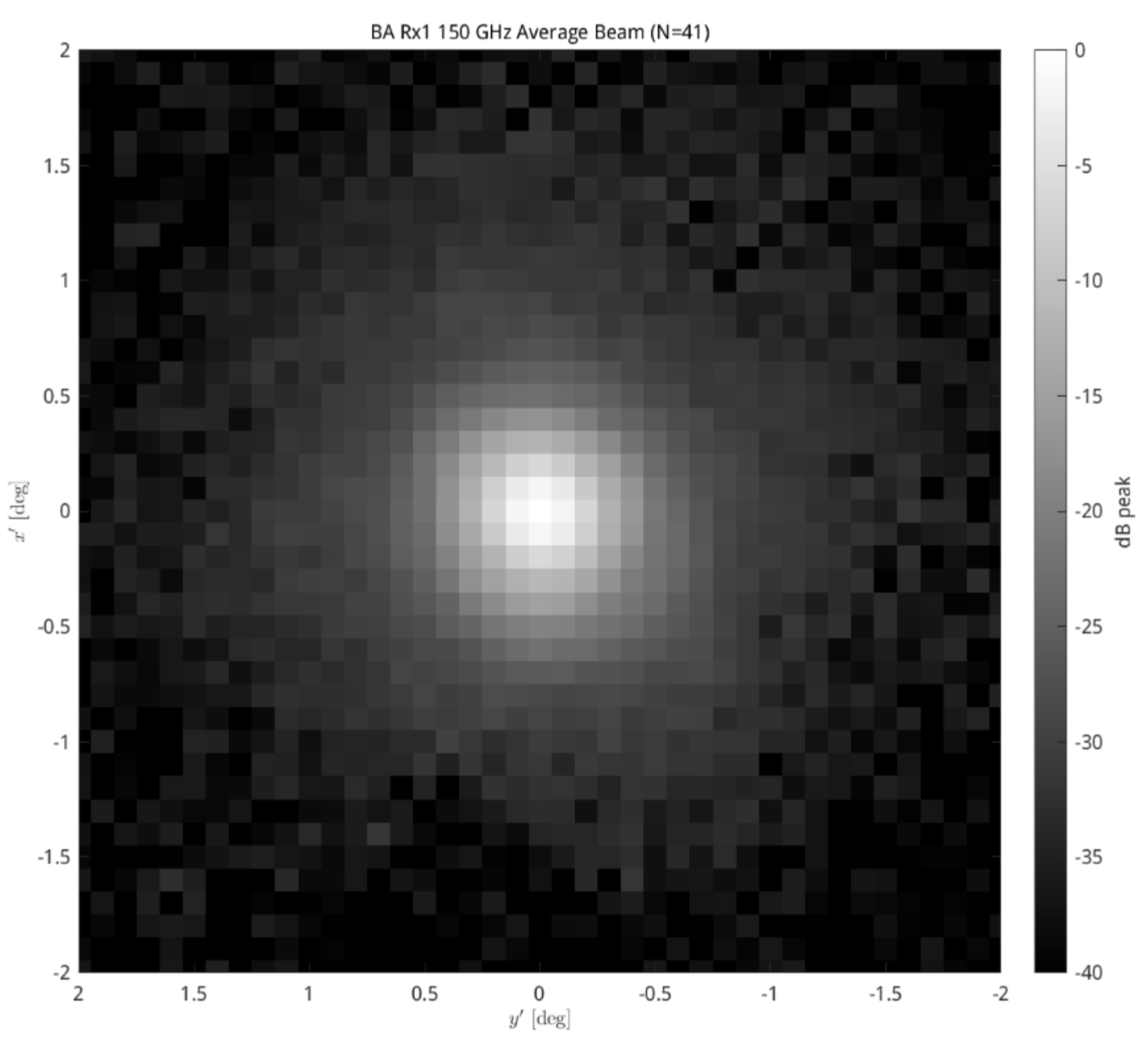}
    \caption{BA2-150 average beam, made by coadding composite beam maps from 41 hand-picked composites.}
    \label{fig:compbeam}
\end{figure}

High-signal-to-noise array-averaged beam maps are constructed by co-adding individual detector composite maps after hand-selecting detectors for data quality and pointing alignment. Figure~\ref{fig:compbeam} shows the resulting best-estimate average beam at 150~GHz, coadded from 41 high-quality composites using the component maps from the 2026 far field beam map campaign.

\section{BEAM MAP PIPELINE IMPROVEMENTS}
\label{sec:sections}

\begin{figure}%[htbp]
    \centering
    \includegraphics[width=0.9\linewidth]{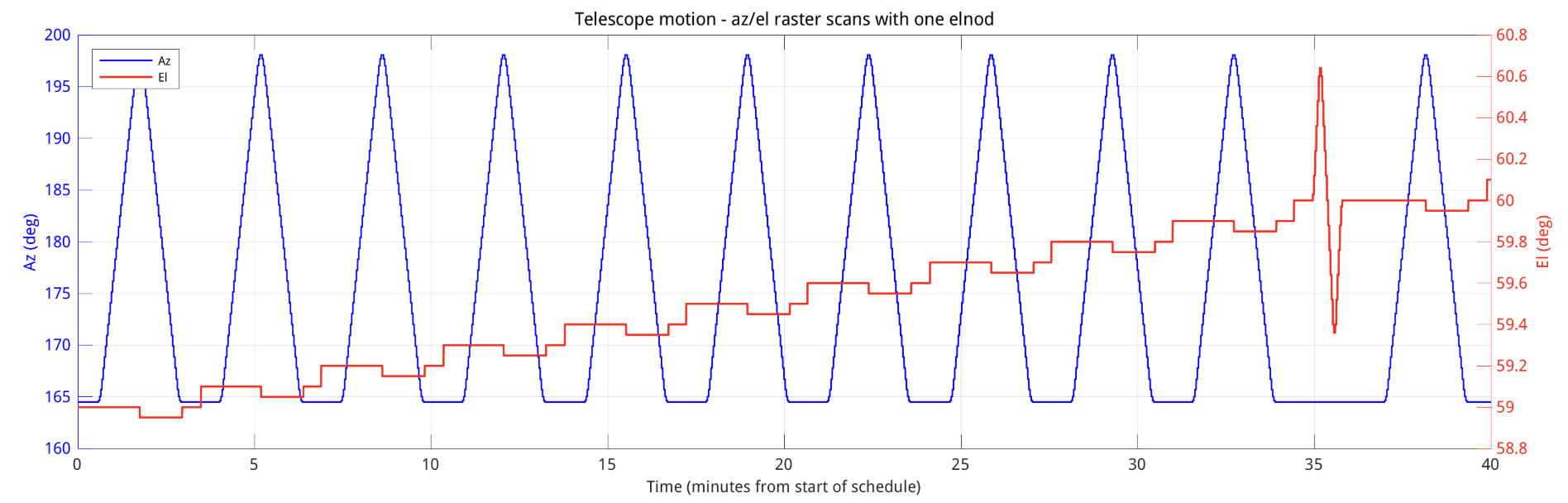}
    \caption{Azimuth (blue) and elevation (red) of raster scans from 40 minutes of the FFBM schedule. Each sweep corresponds to one azimuth pass but the two legs are not at the exact same elevation, the return leg is offset by -0.05° to the outbound one, the outer loop then steps up by 0.15°}
    \label{fig:az}
\end{figure}

The telescope trajectory during a typical beam mapping schedule is shown in Fig.~\ref{fig:az}; we take azimuth scans at 0.05° elevation steps. Beam maps are typically binned in pixels of 0.1°, providing two passes per pixel. This also allows us to perform scan-direction jackknives (null tests) within the 0.1° pixel, removing common-mode sky and source signal while providing an estimate of statistical noise in the beam maps. Analysis of these jackknives is ongoing.

%To better quantify the noise in the beam maps, we implemented an out-and-back jackknife using back-and-forth scans in azimuth that make up each FFBM schedule. Each scan command in the schedule executes a full right to left azimuth sweep as a single unit; differencing these two legs isolates the noise contribution while common-mode sky and source signal cancel. The two legs are not at the exact same elevation, the outbound leg runs at elevation $el_0$, while the return leg is offset by -0.05° to $el_0$, the outer loop then steps $el_0$ up by 0.1°. 

We also investigated an elevation nod (``elnod'')-based approach to gain calibration for the FFBM pipeline. An elnod injects a small, known amount of atmospheric signal into the detector timestreams via a brief, deliberate elevation excursion, providing an independent, per-detector gain reference distinct from the peak-normalization gain delivered from the Gaussian beam fit itself. 

Figure~\ref{fig:az} shows both of these together over a representative 40 minute window of an FFBM schedule from the 2026 campaign. Azimuth traces out the back-and-forth raster (blue), while elevation (red) shows the staircase of 0.05° steps taken between successive raster passes, rising from about 59.0° to 60.1° over the course of the window as the schedule steps up. Partway through, at around minute 35, a single elnod interrupts the climb. 

\section{CONCLUSIONS}
In  these proceedings we have presented a preliminary analysis of the 2026 far field beam map campaign for BICEP Array. This campaign prioritized receiver health and performance checks for the newly commissioned BA3-220/270 receiver, and extended BA2-150 coverage to the newly installed tiles. After correcting from convolution with the finite chopper aperture, we confirmed the expected beamwidth ordering of $\sigma_{150} > \sigma_{220} > \sigma_{270}$ across the mapped bands, and confirmed general consistency in BA2-150 before and after 2024--2026 edge-filter change.

Advances in the data-taking and pipeline include an out-and-back jackknife of the azimuth raster scans, providing a handle on the FFBM noise budget, and an elnod-based gain calibration as an alternative, independent gain reference to the standard peak normalization approach. The next steps are to extend the elnod-based calibration to the full FFBM pipeline, and use the calculated differential beam parameters and higher-order residual beams in the 2026 temperature-to-polarization leakage analysis.

\acknowledgments 
The BICEP/$Keck$ experiments have been funded through U.S. National Science Foundation grants most recently including 2220444-2220448, 2216223, 1836010, and 1726917. The research carried out at the Jet Propulsion Laboratory, California Institute of Technology, under a contract with the National Aeronautics and Space Administration (80NM0018D0004). Focal plane development and testing were supported by the Gordon
and Betty Moore Foundation at the California Institute of Technology. Readout electronics were supported by the Canada Foundation for Innovation grant to the University of British Columbia. The computations in this paper were run on the Cannon cluster supported by the FAS Science Division Research Computing Group at Harvard University. The analysis effort at Stanford University and the SLAC National Accelerator Laboratory was partially supported by the Department of Energy. We thank the staff of the U.S. Antarctic Program and in particular the South Pole Station without whose help this research would not have been possible. We also thank our winter-over operators: Manwei Chan, Karsten Look, Calvin Tsai, Paula Crock, Ta Lee Shue, Grantland Hall, Hans Boenish, Robert Schwarz, Sam Harrison, Anthony DeCicco, Thomas Leps, Brandon Amat, Nathan Precup, Steffen Richter, Thibault Romand, Danielle Simmons, Markus Ayasse, Steven Jungst, Nathan McReynolds, and John Della Costa.

% References

\end{document}